\documentclass[fleqn,twoside]{article}
\usepackage{espcrc2}
\usepackage{epsfig}

\title{ Ultra high energy cosmic rays: clustering, GUT scale and neutrino
masses}

\author{Z. Fodor\address{
Institute for Theoretical Physics, E\"otv\"os University,
P\'azm\'any 1, H-1117 Budapest, Hungary}}

\begin{document}

\begin{abstract}
The clustering of ultra high energy (above $5\cdot 10^{19}$~eV) 
cosmic rays (UHECR) suggests that they might be emitted by compact 
sources.  We present a statistical analysis on the source 
density based on the multiplicities. The propagation of UHECR protons 
is studied in detail.  The UHECR spectrum is consistent with the 
decay of GUT scale particles and/or with the Z-burst. The 
predicted GUT mass is $m_X=10^b$~GeV, where $b=14.6_{-1.7}^{+1.6}$.  
Our neutrino mass prediction depends on the origin of the power 
part of the spectrum: $m_\nu=2.75^{+1.28}_{-0.97}$ eV for 
halo and $0.26^{+0.20}_{-0.14}$~eV for extragalactic (EG) 
origin. 
\end{abstract}

\maketitle

\section{Introduction}
The interaction of protons with the microwave 
background predicts a drop in the cosmic ray flux above
the GZK \cite{GZK66} cutoff  $\approx$5$\cdot$10$^{19}$~eV. The 
data shows no such drop. 
Section 2 studies the
propagation and determines the probability $P(r,E,E_c)$
that protons created at distance $r$ with
energy $E$  reach earth above a threshold $E_c$. Using this $P$
one can give the observed spectrum by one numerical integration
for any injection spectrum.

It is an interesting phenomenon that the UHECR
events are clustered.
Recently, a statistical analysis \cite{DTT00} based on the multiplicities
of the clustered events estimated the source density.
In Section 3 we extend the above analysis. 

In Section 4 we study the scenario that the UHECRs are coming
from decaying superheavy particles (SP) and we determine their masses  
$m_X$ by an analysis of the observed UHECR spectrum. 

Ultrahigh energy neutrinos (UHE$\nu$) scatter on 
relic neutrinos (R$\nu$) producing Z bosons, which can decay hadronically 
(Z-burst) \cite{FMS99}. 
In Section 5 we compare the  predicted proton spectrum with 
observations and determine the mass of the heaviest R$\nu$ via a 
maximum likelihood analysis. 

The details of the presented results and a more complete reference list 
can be found in \cite{FK00,FK01,FKR01}. 

\section{Propagation of UHECR protons}

\begin{figure}\begin{center}
\epsfig{file=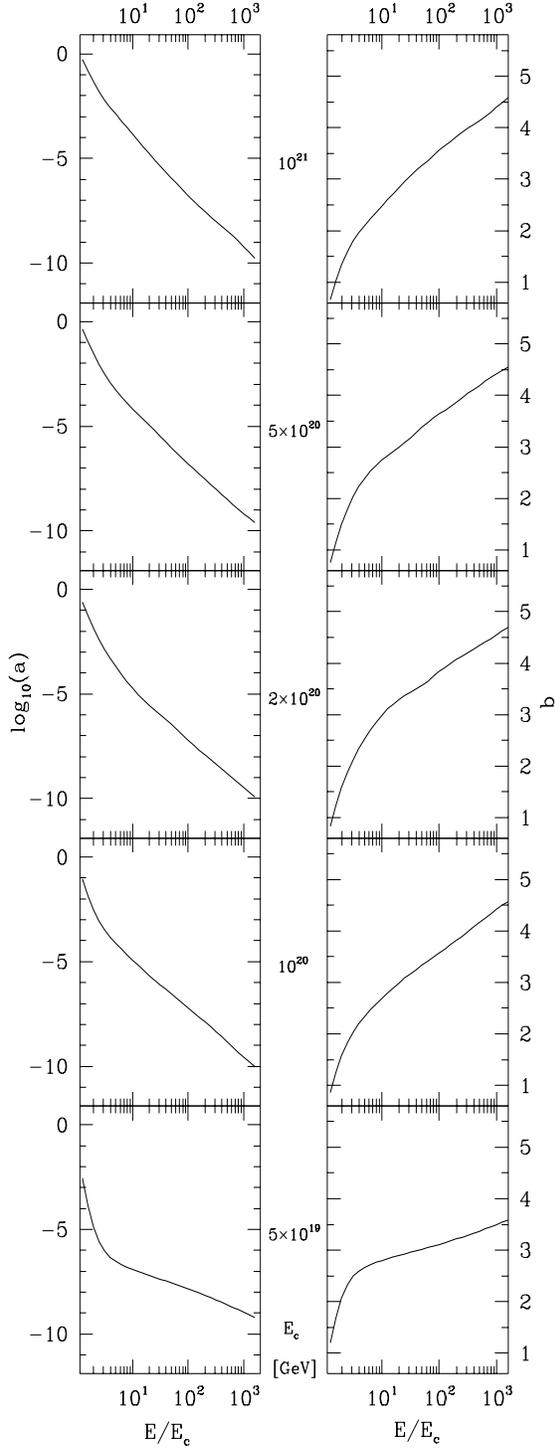,width=8.5cm,height=18.5cm,bbllx=210,bblly=200,bburx=460,bbury=700}
\caption{\label{gzk}
{  
The parametrization of $P(r,E,E_c)$.
}}
\end{center}\end{figure}

Using pion production as the dominant effect of energy loss for
protons at energies $>$10$^{19}$~eV ref. \cite{BW99} calculated 
$P(r,E,E_c)$ for three threshold energies. 
We extended the results of \cite{BW99}. 
In our Monte-Carlo approach protons are propagated in small steps
($10$~kpc), and after each step the probabilities of pion production
and the energy losses due to pair
production, pion production and the adiabatic expansion are calculated.
We used the following type of parametrization
$P(r,E,E_c)=\exp\left[ -a\cdot(r/1\ {\rm Mpc})^b\right]$.
Fig. \ref{gzk} shows  $a(E/E_c)$ and $b(E/E_c)$
for a range of three orders of magnitude and for five different
$E_c$. 

\section{Density of sources}

The arrival directions of
the UHECRs measured by experiments show some peculiar clustering:
some events are grouped within $ \sim 3^o$, the typical angular
resolution of an experiment. Above $4\cdot 10^{19}$ eV 92 cosmic ray events
were detected, including 7 doublets and 2 triplets.
Above $10^{20}$ eV one doublet out of 14 events were found \cite{Uchi}.
The chance probability of such a clustering from uniform distribution is
rather small \cite{Uchi,Hea96}. 

The clustered features of the events initiated
an interesting statistical analysis
assuming compact UHECR sources \cite{DTT00}. The authors found
a large number, $\sim 400$ for the number of
sources within the GZK sphere. We generalize their analysis.
The most probable value for the source density is really large; 
however, the statistical significance of this result is rather weak. 

\begin{figure}\begin{center}
\epsfig{file=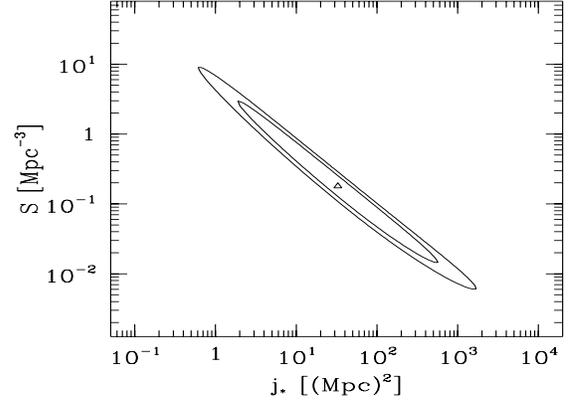,width=7.0cm,height=4.5cm,bbllx=30,bblly=250,bburx=550,bbury=700}
\caption{\label{ell}
{ The $1\sigma$ (68\%) and $2\sigma$ (95\%) confidence
level (CL) regions for $j_*$ and the
source density (14 UHECR with one doublet). 
}}
\end{center}\end{figure}

Fig. \ref{ell} shows the CL regions for one of our
models (injected energy distribution
$c(E) \propto E^{-3}$; 
Schechter's luminosity distribution: 
$h(j)dj \propto (j/j_*)^{-1.25}\exp(-j/j_*)d(j/j_*)$).
The regions are deformed, thin ellipse-like objects.
For this model our final answer for the density is
$180_{-165(174)}^{+2730(8817)}\cdot 10^{-3}$~Mpc$^{-3}$,
where the first errors
indicate the 68\%, the second ones in the parenthesis the 95\%
CLs, respectively.
The choice of \cite{DTT00} --h(j)$\propto\delta$(j)--
and, e.g. $E^{-2}$ energy distribution
gives much smaller value:
$2.77_{-2.53(2.70)}^{+96.1(916)} 10^{-3}$~Mpc$^{-3}$, which is in a quite
good agreement with their result.

\section{Decay of GUT scale particles}

\begin{figure}\begin{center}
\epsfig{file=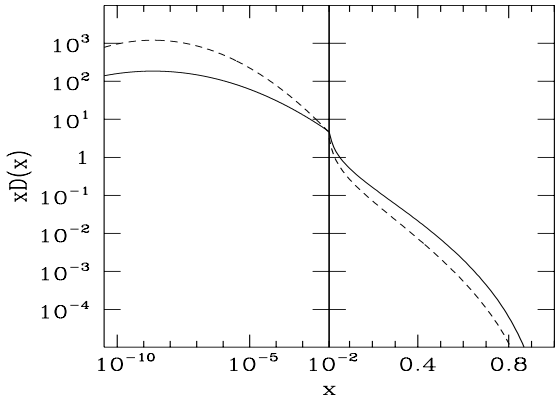,width=7.0cm,height=4.5cm,bbllx=30,bblly=260,bburx=550,bbury=700}
\caption{\label{fragmentation}
{The quark FFs 
at Q=10$^{16}$ GeV for proton/pion in SM (solid/dotted line)  and in MSSM 
(dashed/dashed-dotted line). 
We change from logarithmic scale to linear at $x=0.01$.
}}
\end{center}\end{figure}

An interesting idea discussed by refs.\cite{BKV97,KR98,BS98} is that 
SPs could be the source of UHECRs. 
The hadronic decay of SPs yields protons. They are characterized 
by the fragmentation function (FF) $D(x,Q^2)$ which gives the number of produced 
protons with momentum fraction $x$ at energy scale $Q$.
For the proton's FF at present accelerator
energies we use ref. \cite{BKK95}. We evolve
the FFs in ordinary and 
in supersymmetric QCD to the energies of the 
SPs. This result can be combined with the
prediction of the MLLA technique, which gives 
the initial spectrum of UHECRs at the energy $m_X$ (cf. Fig.. 
\ref{fragmentation}). Similar results are obtained by
\cite{ST01}. 

Depending on the location of the
source --halo or extragalactic (EG)-- and the 
model --SM or MSSM-- we study four different scenarios. In the EG case 
protons loose some fraction of their energies, described
by $P(r,E,E_c)$. We compare the predicted and
the observed spectrums by a maximum likelihood analysis.
This analysis gives the mass of the SP and the error on it.

\begin{figure}\begin{center}
\epsfig{file=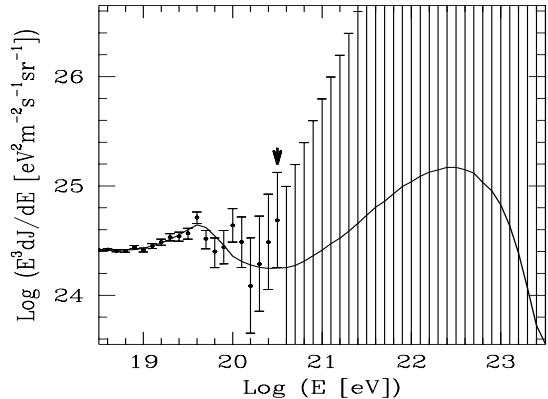,width=7.0cm,height=4.5cm,bbllx=30,bblly=260,bburx=550,bbury=700}
\caption{\label{spect}
{UHECR data with their error bars
and the best fit from a decaying SP.
There are no events above $3 \times 10^{20}$~eV 
(shown by an arrow). 
Zero event
does not mean zero flux, but an upper bound for the flux. 
thus, the experimental 
flux is in the ''hatched'' region with 68\% CL. 
}}
\end{center}\end{figure}

Fig. \ref{spect} shows the measured UHECR spectrum and the best fit, which
is obtained in the EG-MSSM scenario. The
goodnesses of the fits for the halo models are far worse. 
The SM and MSSM cases do not differ significantly. 
The most important message is that the masses of the best fits 
(EG cases) are compatible within the
error bars with the MSSM gauge coupling unification GUT scale: 
$m_X=10^b$~GeV, where $b=14.6_{-1.7}^{+1.6}$.

\section{Z-burst scenario}

Already in the early 80's 
there were discussions that the  UHE$\nu$ spectrum
could have absorption dips at energies around
$E_{\nu_i}^{\rm res}$=$M_Z^2/(2\,m_{\nu_i})$=$4.2\cdot 10^{21}$ 
(1 eV/$m_{\nu_i}$) eV 
due to resonant annihilation with R$\nu$s of mass $m_\nu$, 
predicted by the hot Big Bang, 
into Z bosons of mass $M_Z$~\cite{W82,Y97}. 
Recently it was realized that the same annihilation mechanism gives 
a possible solution to the GZK problem~\cite{FMS99}. 
It was argued that the UHECRs above the GZK cutoff are 
from these Z-bursts.  

We compare this scenario with observations.

The density distribution of R$\nu$s as hot DM follows the total mass
distribution; however, it is less clustered.
Thus we, as opposed to practically all previous
authors~\cite{FMS99,Y99,W98}, do not follow the unnatural assumption of having
a relative overdensity of $10^2\div 10^4$ in our neighborhood (for an approach
with lepton asymmetry see \cite{GK99}).

We give the energy distribution of the produced protons in our lab system, which
is obtained by Lorentz transforming the CM collider results. 

The next ingredient is the propagation of the protons,  which can be
described by $P(r,E_p,E)$.

Finally, we compare the predicted and observed spectrum and 
extract the mass of the R$\nu$ and the necessary UHE$\nu$ flux 
by a maximum
likelihood analysis. In the Z-burst scenario small R$\nu$ mass needs large 
$E_\nu^{\rm res}$ to produce a Z. Large $E_\nu^{\rm res}$ results in 
a large Lorentz boost, thus large proton energy. In this way the
detected energy determines the mass of the R$\nu$. The analysis is completely
analogous to that of the previous section.

Our best fits to the observed data gives for the neutrino mass
$2.75^{+1.28(3.15)}_{-0.97(1.89)}$~eV for the ``halo''-
and $0.26^{+0.20(0.50)}_{-0.14(0.22)}$~eV
for the ``EG''-case, respectively (in the halo/EG cases the power part of the
UHECR spectrum is generated in the halo/EG). 
This gives an absolute lower bound on the mass of the 
heaviest $\nu$ of $0.06$ eV at the 95\% CL. 

The most attractive pattern for $\nu$ masses is hierarchical. Using the 
mass difference of the atmospheric $\nu$ oscillation for the
heaviest mass~\cite{PDG}, one obtains values between 0.03 and 0.09~eV.  
It is an intriguing feature of our result that the smaller
one of the predicted masses is compatible on the $\approx$ 1.3\,$\sigma$ level 
with this scenario. 

\begin{figure}
\begin{center}
\epsfig{file=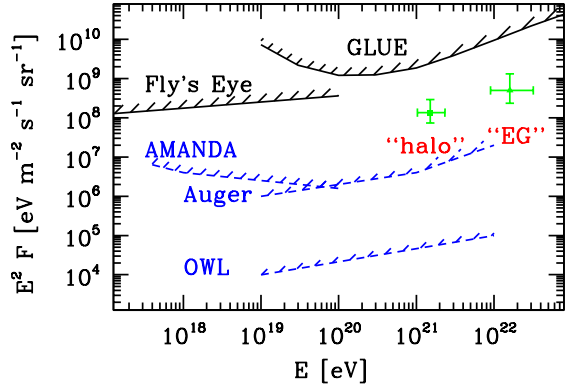,bbllx=20pt,bblly=270pt,bburx=570pt,bbury=608pt,%
width=7.5cm}
\caption[...]{\label{eflux}
Differential $\nu$+$\bar \nu$ fluxes (averaged over the families)
required by the Z-burst hypothesis. 
The horizontal errors indicate the uncertainty of the
mass determination and the vertical errors include also the uncertainty
of the Hubble expansion rate.
Also shown are upper limits from Fly's Eye and the 
GLUE, 
as well as projected sen\-si\-tivi\-ties of AMAN\-DA, 
Auger and OWL.} 
\end{center}
\end{figure}

The necessary UHE$\nu$ flux can be obtained from our fits. 
We summarized them in Fig.~\ref{eflux}, together with some
upper limits.
We also compared our $\gamma$-flux with the EGRET bound~\cite{EGRET}. 
Our $\gamma$ flux is somewhat smaller than that of EGRET.

The nice collaboration with S.D. Katz and A. Ringwald and the careful
reading of the manuscript is acknowledged. 
This work was partially supported by Hung. Sci.
grants No. 
OTKA-\-T34980/\-T29803/\-T22929/\-M28413/\-OM-MU-708/\-IKTA111/\-NIIF.

\end{document}